\documentclass[aps,twocolumn,groupedaddress,showpacs]{revtex4}
\usepackage[dvips]{graphicx}
\usepackage[]{caption}
\usepackage{amsmath}
\usepackage{amssymb}
\def\beq{\begin{equation}}
\def\eeq{\end{equation}}
\def\beqa{\begin{eqnarray}}
\def\eeqa{\end{eqnarray}}
\def\beqan{\begin{eqnarray*}}
\def\eeqan{\end{eqnarray*}}

\begin{document}

\date{\today}
\preprint{Brown-HET-1471}
\title{Predictions of Dynamically Emerging Brane Inflation Models}

\author{Thorsten Battefeld$^{1)}$} \email[email: ]{battefeld@physics.brown.edu}
\author{Natalia Shuhmaher$^{2)}$} \email[email: ]{shuhmaher@hep.physics.mcgill.ca}

\affiliation{1) Dept.of Physics, Brown University, Providence R.I.
02912, U.S.A.} \affiliation{2) Dept.of Physics, McGill University,
Montr\'eal QC, H3A 2T8, Canada}

\begin{abstract}
We confront the recent proposal of Emerging Brane Inflation with
WMAP3+SDSS, finding a scalar spectral index of
$n_s=0.9659^{+0.0049}_{-0.0052}$ in excellent agreement with
observations. The proposal incorporates a
preceding phase of isotropic, non accelerated expansion in all
dimensions, providing suitable initial conditions for inflation.
Additional observational constraints on the parameters of the model
provide an estimate of the string scale.

A graceful exit to inflation and stabilization of extra dimensions
is achieved via a string gas. The resulting pre-heating phase shows
some novel features due to a redshifting potential, comparable to
effects due to the expansion of the universe itself. However, the
model at hand suffers from either a potential over-production of
relics after inflation or insufficient stabilization at late times.

\end{abstract}

\maketitle

%%%%%%%%%%%%%%%%%%%%%%%%%%%%%%%%%%%%
\section{Introduction}
%%%%%%%%%%%%%%%%%%%%%%%%%%%%%%%%%%%%

Inflation provides a natural explanation for major problems of
standard cosmology such as the homogeneity, horizon and flatness
problems~\cite{Guth}. An almost
scale-invariant spectrum of adiabatic cosmological fluctuations was predicted \cite{Lukash,ChibMukh} more
than a decade before the cosmic microwave background anisotropies
were analyzed ~\cite{WMAP,Maxima,Boomerang,COBE}. Encouraged by
the great success of the inflationary paradigm, one is urged to find a
successful realization of inflation within more fundamental theories, such as string theory.

The heuristic approach of obtaining inflation consists of
introducing one or more scalar fields (inflatons), which evolve
slowly due to some appropriately tuned potential. Simple single
field models require the inflaton to start out at a value larger
than the Planck mass. However, at these values radiative corrections
to the inflaton mass threaten to spoil slow roll dynamics.
Therefore, unless some underlying symmetry protects the inflaton
mass, it is hard to implement an inflationary scenario within gauge
theories. Moreover, in the framework of four dimensional
inflationary models the physical interpretation of the inflaton is
unclear. Generally, it is taken to be a singlet of the Standard
Model and typically of all of the visible sector too.
(See~\cite{Allahverdi:2006iq} for a recently proposed exception.)

The advent of  extra dimensions~\cite{ADD,RS} opened up a venue for
new inflationary scenarios where the inflaton has a physical
meaning;  for example, in brane-antibrane inflationary models the
inter-brane separation serves as the inflaton~\cite{DT,Shafi}.
However, to provide a sufficient amount of inflation in
brane-antibrane models, one requires fine tuning of initial
conditions~\cite{Cliff,Cliff2}, e.g. large inter brane separation,
special configurations or very weak couplings.

In this article, we would like to discuss the recent proposal of emerging brane inflation
\cite{size,Shuhmaher:2005pw} that makes use of extra dimensions, a
gas of p-branes in the bulk to drive an initial isotropic but non
accelerated expansion of the universe, as well as orbifold fixed
planes responsible for contraction of the extra dimensions and
inflation of our three dimensions. The presence of a string gas at
the end of inflation provides a graceful exit, pre-heating and
stabilization of extra dimensions. The model at hand does not need
any fine tuning of initial conditions and will turn out to be in
good agreement with observations.

In this scenario the multidimensional universe starts out small and
hot, with our three dimensions compactified on a torus and the extra
dimensions on an orbifold of the same size. The pre-inflationary
expansion \cite{size,Shuhmaher:2005pw} is governed by topological
defects (p-branes) in the bulk and responsible for a large inter
brane separation. As the universe expands isotropically due to the
gas of p-branes, the energy density stored in the gas gets diluted
until additional weak forces come into play, changing the overall
dynamics. For example, branes pinned to orbifold fixed planes, which
exhibit an attractive force, may eventually cause a contraction of
the extra dimensions while our dimensions inflate. From the four
dimensional point of view, the inflaton is identified with the
radion and consequently, the pre-inflationary bulk expansion
explains the large initial value of the inflaton. Inflation comes to
an end when the extra dimensions shrink down to a small scale where
moduli trapping \cite{Watson:2003gf,Patil:2004zp,Kofman:2004yc,Watson:2004aq}
and pre-heating \cite{Traschen:1990sw,Shtanov:1994ce,Kofman:1994rk,Kofman:1997yn,Greene:1997fu,Felder:2000hj} can occur.

Our main goal in this article is to examine the viability of the
emerging brane inflation model outlined above and to  make contact
with observations.

The outline of this article is as follows: In Ch. \ref{model} we
review the details of the model, followed by a computation (Ch.
\ref{predictions}) within the slow roll approximation of the
spectral index $n_s$, the running of the index $d\,n_s/(d\,\ln k)$,
the scalar to tensor ratio $r$ and the tensor index $n_T$. We
confront our predictions with the observation of the cosmic
microwave background radiation measured by the Wilkinson Microwave
Anisotropy Probe  (WMAP3)
\cite{Spergel:2006hy,Page:2006hz,Hinshaw:2006ia,Jarosik:2006ib} and
the Sloan Digital Sky Survey (SDSS) \cite{Tegmark:2001jh}, resulting
in good agreement. In order to get sufficient initial expansion one
requires the inter brane potential to remain subdominant for a long
time compared to the energy density stored in the bulk p-branes.
This requirement imposes constraints on the scale of interactions
between branes pinned to the orbifold fixed points. This, together
with constraints from observational data, will be sufficient to
provide an estimate of the string scale. In Chapter
\ref{preheating}, we study the viability of pre-heating after
inflation and stabilization at late times. While pre-heating can
occur in the standard manner, albeit some novel effects are present,
we do find potential problems associated with either late time
stabilization or relics: if the branes pinned to the orbifold fixed
planes annihilate after inflation they could produce an
over-abundance of relics such as cosmic strings, and if they do not
annihilate they will destabilize the extra dimensions at late times.
We conclude with a comment on open issues within the framework of
emerging brane inflation.

\section{The Model \label{model}}
Following \cite{Shuhmaher:2005pw}, we assume a spacetime
\begin{equation}
\label{orbi} {\cal M}  =  {\cal R} \times T^3 \times T^d /
Z_2 \, ,
\end{equation}
so that our three dimensions have the topology of a torus $T^3$, and
the $d$ extra dimensions are compactified on the orbifold $T^d/Z_2$.
Note, that this specific choice of the
manifold is not crucial for the model -- many other manifolds distinguishing our
three dimensions could be chosen instead. Next, we assume that pairs
of branes are pinned to the different orbifold fixed planes a
distance $r$ apart. Furthermore, we also assume inter brane interactions
such that a weak attractive force is generated via some potential
$V$. It should be noted that we take a phenomenological approach and
postulate the existence of a potential with the desired properties.
A discussion of possible origins of inter-brane potentials can be
found in \cite{DT}.

The special feature of the underlying scenario is the
pre-inflationary dynamics which explains the large size of the extra
dimensions just before inflation. Initially, the universe starts out
small and hot with all spatial dimensions of the same size. The bulk
is filled with a gas of p-branes. In this phase, the energy density
in the brane gas is assumed to be many orders of magnitude larger
than the potential energy density, which provides the force between
the orbifold fixed planes. Thus, the universe expands
isotropically but not inflationary, as shown in \cite{size}. During
the expansion phase the bulk energy density of the gas decreases
and eventually the potential $V$ begins to dominate, causing
inflation of the directions parallel to the orbifold fixed planes,
and contraction of the extra dimensions. It should be noted that
whether or not inflation occurs is sensitive to the form of the
inter-brane potential $V$.

Following~\cite{Shuhmaher:2005pw}, we consider a potential of the
form
\begin{equation}
\label{pot} V(r)  =  - \mu {1 \over r^n} \, ,
\end{equation}
where $r$ is the inter brane separation and $n>0$ is a free parameter, which could in principle be computed
from the underlying fundamental theory. As we shall see below, this form of the potential results in inflation.

We shall first compute how dynamics can be described in a four
dimensional effective theory. Let $G_{ab}$ be the metric for the
full space-time with coordinates $X^a$. In the absence of spatial
curvature, the metric of a maximally symmetric space which
distinguishes 'our' three dimensions is given by
\begin{equation}
ds^2  =  G_{ab} dX^a dX^b  \, = \, d{t'}^2 - \alpha(t')^2 d{\bf x}^2
- b(t')^2 d{\bf y}^2 \, ,
\end{equation}
where ${\bf x}$ denotes the three coordinates parallel to the
orbifold fixed planes and ${\bf y}$ denotes the coordinates of the $d$
perpendicular directions.

Our goal is to find a four-dimensional effective potential which
governs the inflationary phase. We start out with the higher dimensional
action
\begin{equation}
S  =  \int d^{d+4}X \sqrt{-\det G_{ab}} \left\{ \frac{1}{16
\pi G_{d+4}} R_{d+4} + \hat{\cal L}_M \right\} \,,
\end{equation}
where $R_{d+4}$ is the $d+4$ dimensional Ricci scalar and $\hat{\cal
L}_M$ is the matter Lagrangian density with the metric determinant
factored out. Note that the dilaton is assumed to be fixed already,
e.g. via the proposal of \cite{Cremonini:2006sx}. In the effective
four-dimensional action, the radion $b(t)$ is replaced by a
canonically normalized scalar field $\varphi(t)$ which is related to
$b(t)$ through~\cite{Battefeld:2005av,Carroll:2001ih}
\begin{equation}
\label{rel} \varphi  = \beta^{-1} m_{pl} \ln (b) \, ,
\end{equation}
where
\begin{equation}
m^2_{pl} = \frac{1}{8 \pi G_4}
\end{equation}
is the reduced four dimensional Planck
mass and we defined
\begin{equation}
 \beta^{-1} :=
{\sqrt{d(d+2) \over 2}} \,.
\end{equation}
After performing a dimensional reduction and a conformal
transformation to arrive at the Einstein frame
\cite{Battefeld:2005av,Carroll:2001ih} we are left with
\begin{eqnarray}
S  =  \int d^4 x a^3 \left\{ \frac{1}{2} m_{pl} R_4
\right. &-& \left. \frac{1}{2} (\partial \varphi)^2 \right. \\ &+&
\left. {\cal V} e^{- d \varphi/m_{pl}\beta} \hat{\cal L}_M \right\}\,,
\nonumber
\end{eqnarray}
where
\begin{equation}
{\cal V} \, = \, \int d^d{\bf y} \, = \, l^d_s
\end{equation}
is the volume of the extra dimensions, and
\begin{eqnarray}
\nonumber ds_E^2 &=&  b^d (dt^{\prime 2} - \alpha^2 d{\bf x}^2)\\
 &=&  dt^2 - a(t)^2 d{\bf x}^2
\end{eqnarray}
is the effective four dimension metric. Further, assuming that the
initial separation between the orbifold fixed planes is of string
length $l_s$ one can compute the distance between the orbifold fixed
planes to
\begin{equation}
r(t')  =  l_s b(t') \, .
\end{equation}
Note that $b=1$ corresponds to the string scale. Setting $\hat{\mathcal{L}}_M=V$ yields
\begin{eqnarray}
\label{pot2} V_{4d}(\varphi) \, &=& \, g^4_s l^d_s b(\varphi)^{-d} V(r(\varphi))\nonumber \\
 &=& -\mu g^4_s l^{d-n}_s e^{-\tilde \alpha {\varphi \over {m_{pl}}}} \, ,
\end{eqnarray}
where we used (\ref{pot}), restored the string coupling dependence and defined
 \begin{equation}
{\tilde \alpha} := (n+d) \beta \,.
\end{equation}
To account for the brane tension/zero cosmological constant today we
add a positive constant $V_0$ to the effective potential
(\ref{pot2}) and arrive at the effective four dimensional potential
\begin{eqnarray}
 V_{eff}(\varphi)  &=&  V_0 -\mu g^4_s l^{d-n}_s e^{-\tilde \alpha {\varphi \over {m_{pl}}}} \nonumber \\
&=& V_0 (1- \zeta e^{- \tilde \alpha/m_p \, \varphi})  \label{effpot}
\end{eqnarray}
where we defined
\begin{equation}
\label{zeta} \zeta := {\mu g^4_s l^{d-n}_s \over V_0} \, .
\end{equation}

This potential yields inflation for large enough values of the
radion/inflaton $\varphi$. Similar potentials have been considered
before, see e.g.~\cite{Conlon:2005jm} %\footnote{We thank C. Burgess
%for pointing out the reference~\cite{Conlon:2005jm}.}
 in the context
of brane inflation or \cite{Stewart:1994ts} in the context of
supergravity. These proposals differ from ours in the form of the
graceful exit, the details of pre-heating as well as the
 the pre-inflationary dynamics of our
model, which pull $\varphi$ far from its minimum such as to provide
suitable initial conditions for inflation without fine tuning.

\section{Predictions \label{predictions}}
Inflation gives rise to a viable mechanism of structure formation:
quantum vacuum fluctuations, present during inflation on microscopic
scales, exit the Hubble radius and are subsequently squeezed,
resulting in classical perturbations at late times, see
e.g.~\cite{Mukhanov:1990me}. Moreover, current observations are
precise enough to distinguish between different inflationary models
\cite{Spergel:2006hy,Page:2006hz,Hinshaw:2006ia,Jarosik:2006ib,Tegmark:2001jh,Kinney:2006qm,Martin:2006rs}.

In the following, we derive observable quantities within the slow
roll approximation and make contact with observations. Thereafter,
we show how one can estimate the string scale in the model at hand.

\subsection{Cosmological Parameters}

The equation of motion for a scalar field in an expanding universe
is given by
\begin{equation} {\ddot \varphi} + 3H \dot \varphi  + V_I^\prime  =  0\,,
\label{EOM}
\end{equation}
where $H = \dot a /a$ is the Hubble parameter, $V_I:=V_{eff}$ from
(\ref{effpot}) is the inflaton potential and $V_I^\prime :=
dV_I/d\phi $. If the scalar field $\varphi$ governs the evolution of
the Universe, the Friedmann Robertson Walker equations become
\begin{eqnarray}
H^2  &=&  \frac{1}{3 m^2_{pl}}
\left[\frac{1}{2} {\dot \varphi}^2 + V_I(\varphi) \right]\,, \label{Hubble} \\
\left(\frac{\ddot a}{a}\right) &=& \frac{1}{3 m^2_{pl}}
\left[V_I(\varphi) - {\dot \varphi}^2 \right] \, .
\end{eqnarray}
If the potential energy of the inflaton dominates
over the kinetic energy, accelerated expansion of the universe results.
In other words, if the potential is flat enough to allow for slow roll of the inflaton field, inflation occurs.
In this case (\ref{EOM}) and (\ref{Hubble}) become
\begin{eqnarray}
3H \dot \varphi &=&- V_I^\prime  \,, \\
H^2 &=& \frac{V_I}{3 m^2_{pl}}\,, \label{slowroll}
\end{eqnarray}
where we assumed ${\dot \varphi}^2 \ll V_I$ and ${\ddot
\varphi} \ll 3H \dot \varphi$.

This approximation is valid if both the slope and the curvature of
the potential are small, that is if the slow roll parameters
\begin{eqnarray}
\varepsilon &=& \frac{m^2_{pl}}{2} \left( \frac{V_I^\prime}{V_I}  \right)^2\,, \label{slp1}\\
\eta &=& m^2_{pl} \frac{V_I^{\prime\prime}}{V}
 \, . \label{slp2}
\end{eqnarray}
satisfy $\varepsilon\ll1$ and $\eta\ll1$. Let $\varphi_i$ denote the
value of the inflaton field N e-folds before the end of inflation.
This value can be determined from
\begin{eqnarray}
\nonumber N &=& \int^{t_{f}}_{t_i} H(t) dt \\
&\approx& \frac{1}{m^2_{pl}}\int^{\varphi_i}_{\varphi_{f}} \frac{V}{V^\prime} d\varphi \,, \label{efoldings}
\end{eqnarray}
where $\varphi_f$ is the value of the inflaton field at which the slow roll approximation breaks down.

Within the slow roll regime one can then compute the scalar spectral
index, the scalar to tensor ratio and the tensor spectral index to
\cite{Kinney:2006qm}
\begin{eqnarray}
n_s&\approx&1-6\varepsilon+2\eta\,,\label{obs1}\\
r&\approx&16\epsilon\,,\\
n_T&\approx&-r/8\label{obs3}\,,
\end{eqnarray}
where $\epsilon$ and $\eta$ have to be evaluated at $\varphi_i$.
For the potential (\ref{effpot}) the slow roll parameters become
\begin{eqnarray}
\varepsilon & = & \frac{\tilde \alpha^2}{2}
\frac{1}{(\zeta^{-1}
e^{\tilde \alpha/m_{pl} \, \varphi} - 1)^2} \, , \\
\eta & = & -\tilde \alpha^2 \frac{1}{\zeta^{-1} e^{\tilde \alpha/m_p
\, \varphi} - 1} \, .
\end{eqnarray}
Since $|\eta| > \varepsilon$ in our case, inflation ends once
$|\eta| = \mathcal{O}(1)$, that is once $\varphi$ approaches
\begin{equation}
\varphi_{f} = \frac{m_{pl}}{\tilde \alpha} \ln{((\tilde \alpha^2 + 1)\zeta)} \, . \label{phifinal}
\end{equation}
By using $V_I$  from (\ref{effpot}) in (\ref{efoldings}) we can
compute the required initial value of the inflaton by solving
\begin{eqnarray}
N &=& \frac{e^{\tilde \alpha/m_{pl} \, \varphi_i} - e^{\tilde
\alpha/m_{pl} \, \varphi_f}}{{\tilde \alpha}^2 \, \zeta} + \, \frac{
(\varphi_f - \varphi_i)}{{\tilde \alpha} \, m_{pl}}\,, \label{integ}
\end{eqnarray}
for $\varphi_i$, which can be done analytically. Neglecting
$\mathcal{O}(1)$ terms in (\ref{phifinal}) gives $\varphi_{f}
\approx \frac{m_{pl}}{\tilde \alpha} \ln{(\tilde{\alpha}^2\zeta)}$
which in turn leads to
\begin{eqnarray}
N&\approx& \, \frac{1}{{\tilde \alpha}^2} \left(e^{\tilde
\alpha/m_{pl} \, \varphi} - \tilde \alpha^2\right) \, ,
\end{eqnarray}
after neglecting the second term in (\ref{integ}). This expression can be solved to
\begin{equation}
\varphi_{i} \approx \frac{m_{pl}}{\tilde \alpha} \ln{(\tilde
\alpha^2 \zeta N + \tilde \alpha^2 \zeta)} \, . \label{varphiN}
\end{equation}
The slow roll parameters (\ref{slp1}) and (\ref{slp2}) evaluated at $\varphi_i$ can now be approximated by
\begin{eqnarray}
\epsilon&\approx&\frac{1}{2\tilde{\alpha}^2\left(N+1 \right)^2}\sim\mathcal{O}\left(\frac{1}{(N\tilde{\alpha})^2}\right)\,,\\
\eta&\approx&-\frac{1}{N+1}\sim\mathcal{O}\left(\frac{1}{N}\right)\,.
\end{eqnarray}
Henceforth, the scalar spectral index becomes
\begin{eqnarray}
n_s&\approx& 1-\frac{2}{N+1}\\
&\approx& 1-\frac{2}{N}\,,
\end{eqnarray}
whereas the scalar to tensor ratio and the tensor spectral index read
\begin{eqnarray}
n_T&=&-\frac{r}{8}\approx-\frac{1}{\tilde{\alpha}^2\left(N+1\right)^2}\\
&\approx&-\frac{1}{\tilde{\alpha}^2N^2}\,.
\end{eqnarray}
In addition, the running of the scalar spectral index can be evaluated to
\begin{eqnarray}
\frac{d\,n_s}{d\,\ln(k)}&=&-m_{pl} \frac{V_I^{\prime}}{V_I}\frac{d\,n_s}{d\,\varphi}\\
&\approx&-\frac{2}{N^2} \,,
\end{eqnarray}
which is negligible.

Now, we can evaluate (\ref{obs1})-(\ref{obs3}) after
specifying some parameters: first, since our model is motivated by
string theory and the dilaton is fixed, we have $d=6$ extra
dimensions. Second, as we shall see in section \ref{preheating},
stabilization of the extra dimensions after inflation at the string
scale is possible. Lastly, we have to specify the
exponent $n$ in (\ref{effpot}), which in turn determines
$\tilde{\alpha}$. This last parameter will barely influence the
scalar spectral index but has some effect on the scalar to tensor
ratio and the tensor spectral index. If we take $n=4$ \footnote{We
have $n=4$ in our specific setup, since $n=d+3-p-2$ with $p=3$ if we
put a 3-brane on the orbifold fixed-planes.} and $N=54\pm 7$ we get
\begin{eqnarray}
n_s&=&0.9659^{+0.0049}_{-0.0052}\,,\\
r&=&\left(6.1^{+1.9}_{-1.3}\right)\times 10^{-4}\,,\\
n_T&=&\left(-7.6^{+2.4}_{-1.6}\right)\times 10^{-5}\,,
\end{eqnarray}
where we used the more cumbersome exact analytic expressions within
the slow roll regime. These predictions \footnote{The limit $p \rightarrow -\infty$ of \cite{Alabidi:2006qa} corresponds to an exponential potential like the one discussed in this article; however, no estimate of $r$ and $n_T$, which depend on the exponent $\tilde{\alpha}$, were given, and the WMAP3 data set alone was used for comparison. This led Alabidi and Lyth to conclude that an exponential potential would be allowed at the $2\sigma$ level.} can now be compared with
observational data. To be specific, the combined observational data of
WMAP3
\cite{Spergel:2006hy,Page:2006hz,Hinshaw:2006ia,Jarosik:2006ib} and
SDSS \cite{Tegmark:2001jh} was used by Kinney et.al. in
\cite{Kinney:2006qm}: the above predictions for $n_s$ and $r$ lie in
the middle of the $1\sigma$ region in the case of negligible running
(see Fig. $1$ in \cite{Kinney:2006qm}).

Hence, the model of emerging brane inflation of
\cite{Shuhmaher:2005pw} passes this first observational test.

\subsection{Estimate of the Fundamental String Length}

In the proposed scenario, brane inflation emerges after the inflaton got pushed up
 its potential in the preceding bulk expansion phase. The
inflaton is related to the scale factor of extra dimensions, $b$,
through~(\ref{rel}). Therefore, the requirement to obtain N
e-foldings of inflation~(\ref{varphiN}) leads to a constraint on the
minimal value of $b$ at the beginning of inflation,
\begin{equation}
b_i \gtrsim (\tilde \alpha^2 N \zeta)^{\beta/\tilde \alpha} \,.
\label{min_b}
\end{equation}

On the other hand, the preceding expansion phase sets an upper limit
on the scale factor $b$~\cite{Shuhmaher:2005pw}. The end of bulk
expansion and the beginning of inflation is indicated
by $V = \rho_b$ where $\rho_b$ is the energy density of the brane gas.
This yields the condition
\begin{eqnarray}
\mu l^{-n}_s b^{-n} &=& l^{-d-4}_s b^{-d-3+p}\,,
\end{eqnarray}
where we assumed that the energy density stored in
p-branes at the beginning of the bulk expansion phase is of the order of
the string scale. Rearranging parameters
leads to
\begin{equation}
 b_i \lesssim (\mu l^{d+4-n}_s)^{-1/(d+3-p-n)} \label{max_b} \, .
\end{equation}
The bound~(\ref{max_b}) relates the scale of the inter-brane
potential $\Lambda = \mu^{1/(d+4-n)}$ to the scale factor of the
extra dimensions. Therefore, the requirement of $N$ e-folds set an
upper bound on $\Lambda$
\begin{equation}
 (\Lambda l_s)^{d+4-n} \lesssim (\tilde \alpha^2 N
\zeta)^{-\beta/\tilde \alpha (d+3-p-n)}\,. \label{bound}
\end{equation}

Next, we can use observational data to constrain the effective inflationary
potential. To be specific, COBE data implies \cite{book} for the scale of inflation
\begin{equation}
 \Big( {V \over \varepsilon} \Big)^{1/4} = 0.027 m_{pl} \, .
\end{equation}
 Evaluating this expression N e-folds before the end of inflation
leads to
\begin{eqnarray}
0.027 m_{pl} &\simeq& V_0 {(\sqrt{2} \tilde \alpha
N)^{1/2}}
\\ &=& {\mu l^{d-n}_s g^4_s (\sqrt{2} \tilde \alpha N)^{1/2} \over
\zeta} \\ &\lesssim& { g^4_s (\sqrt{2} \tilde \alpha N)^{1/2} \over
\zeta \, (\tilde \alpha^2 N \zeta)^{(d+3-p-n)/(d+n)} } \, l^{-4}_s\,,
\end{eqnarray}
where we used (\ref{bound}) in the last expression.
Substituting  $l^{-2}_s$ with $g^2_s m^2_{pl}$, one eventually
arrives at a constraint for the string coupling
\begin{equation}
g^8_s \gtrsim 0.027
\, {\zeta \, (\tilde \alpha^2 N \zeta)^{(d+3-p-n)/(d+n)} \over
(\sqrt{2} \tilde \alpha N)^{1/2}}  \, . \label{string_coupling}
\end{equation}
For $(\zeta, d, p, n, N) = (1, 6, 3, 4, 54)$ this
expression reduces to
\begin{equation}
g_s \gtrsim 0.53\,.
\end{equation}

In conclusion, we found in the model at hand that inflation of about 60 e-folds requires
the string scale to be slightly below the Planck scale.

A word of caution might be in order here:
we assumed the dilaton to be fixed throughout bulk expansion and inflation; but if the dilaton is rolling during these early stages, it will modify the above estimate. Hence, a better understanding of the dilatons stabilization mechanism is of great interest.

\section{Stabilization and Pre-heating \label{preheating}}
We saw in the previous sections how brane inflation can emerge in a
higher dimensional setup. The specific inflaton potential in the
effective four dimensional description was given by (\ref{effpot})
\begin{eqnarray}
V_I=V_0\left(1-e^{-\tilde{\alpha}\varphi}\right)\,, \label{pottb1}
\end{eqnarray}
where we set $m_p\equiv1$ and fine tuned $\zeta=1$. The inflaton is
related to the radion via (\ref{rel}) where $\beta^{-2}=d(d+2)/2$
was introduced. Furthermore, we assumed an already stabilized
dilaton, e.g. via the proposal of \cite{Cremonini:2006sx}. It should
be noted that a free dilaton could potentially invalidate the
predictions of the model at hand.

In the following we would like to address three questions: How does
inflation end, how does the universe reheat and
can the radion/inflaton be stabilized at late times?

\subsection{Stabilization}
Based on the idea of moduli stabilization at points of enhanced
symmetry
\cite{Watson:2003gf,Patil:2004zp,Kofman:2004yc,Watson:2004aq,Patil:2005fi} it was
advocated in \cite{Battefeld:2005wv,Battefeld:2005av} that an
inflationary phase driven by the radion could be terminated by the
production of nearly massless states if the radion comes close to
such a point \footnote{We focus on the overall volume modulus here
-- all other moduli (e.g. complex structure moduli and K\"ahler
moduli) are assumed to be stabilized already. Since it is not always
possible to find points of enhanced symmetry, one can not use the
notion of quantum moduli trapping \cite{Battefeld:2005av} for all of
them.}. To be specific, if we work within heterotic string theory
($d=6$) such a point of enhanced symmetry could be the self dual
radius corresponding to $\varphi=0$. This was already anticipated by
setting $\zeta=1$ so that the potential $V_I$ in (\ref{pottb1})
vanishes at $\varphi=0$ (the self dual radius) \footnote{By choosing
$\zeta=1$ we effectively set the cosmological constant to zero.}.

%%%%%%%%%%%%%%%%%%%%%%%%%%NS-New
The mechanism for stabilizing moduli at points of enhanced symmetry was illustrated in detail
in~\cite{Watson:2004aq,Kofman:2004yc}, and can be implemented in string gas cosmology. 
In the specific toy model of \cite{Watson:2004aq} it was shown that new massless states,
gauge vectors and scalars, appear at the self dual radius. These
states have to be included in the effective four dimensional action, leading
to trapping of the volume modulus: as the radius shrinks down to the
string size, the evolution becomes non-adiabatic and light
states are produced via parametric resonance. Since the coupling of
moduli to vector states is a gauge coupling, one expects
parametric resonance to be efficient. The produced vectors stop to
be massless as the radius shrinks further, generating an effective
potential for the volume modulus. As a consequence, the size of extra
dimensions ceases to shrink. The mechanism of moduli trapping at enhanced
symmetry points (ESP) was discussed more generally
in~\cite{Kofman:2004yc}: the trapping force is proportional to the
number of states that becomes massless at the ESP, since enlarging the amount
of new light degrees of freedom effectively causes an enhanced 
coupling of the moduli. As a consequence of the larger coupling, the
effectiveness of parametric resonance and the trapping effect are enhanced.
Therefore, points with greater symmetry are dynamically preferred.
%%%%%%%%%%%%%%%%%%%%%%%%%%%%%%%%%%NS-endNew

We refer the interested reader to
\cite{Brandenberger:2005fb,Brandenberger:2005nz} for a basic
introduction and to \cite{Battefeld:2005av} for a technical review
of string gas cosmology, and jump into the discussion right after
the string gas got produced.

As mentioned above, the string gas leads to an
effective potential from a four dimensional point of view which is given by \cite{Battefeld:2004xw,Battefeld:2005av}
\begin{eqnarray}
V_S=\frac{\tilde{N}}{a^3}e^{-\frac{d}{2}\beta\varphi}\sqrt{\frac{q^2}{a^2}+\sinh^2(\beta\varphi)}\,, \label{pottb2string}
\end{eqnarray}
where $q$ parameterizes the momentum of the string gas along the
three large dimensions and $\tilde{N}$ is proportional to the number
density of strings. We shall treat both parameters as free ones
\footnote{Both $\tilde{N}$ and $q$ could in principle be computed
via a study of the production mechanism of the sting gas. This
process shares similarities to pre-heating and in fact overlaps with
the early stages of pre-heating. Consequently, pre-heating might be
influenced (see section \ref{pre-heating}).}. Note the novel feature
that the potential redshifts like matter, unlike potentials usually
encountered for scalar fields.

\begin{figure}[tb]
  \includegraphics[width=\columnwidth]{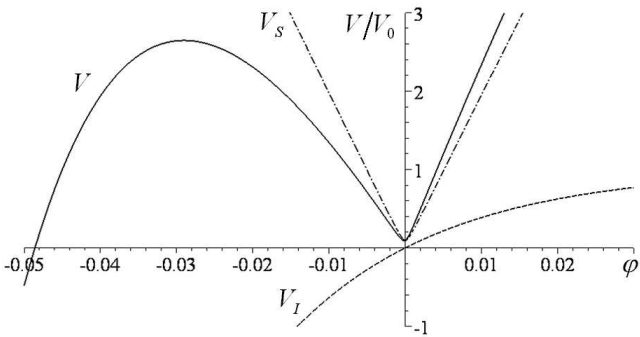}
  \caption{\label{Fig1} The inter-brane potential $V_S$ of
 (\ref{pottb1}), string gas potential $V_I$ of (\ref{pottb2string})
and total potential $V=V_I+V_S$ with $d=6$, $n=4$. The height of $V$
at the minimum $\varphi=\sigma\approx0$ is given by the momentum of
the string gas along the three large dimensions $q$. We chose
$l:=q/a=10^{-6}$ and $\tilde{N}/(V_0a^3)=4\tilde{\alpha}/\beta$ for
instructive reasons only, such that $V_I$, $V_S$ and $V$ are clearly
discernable. Note that $V_S$ is only viable around $\varphi=\sigma$.
In order for a minimum to exist and moduli trapping to occur,
conditions (\ref{cond1stab}) and (\ref{cond2stab}) need to be
satisfied. }
\end{figure}

This redshifting leads to a problem if we insist that the present
day radion be stabilized by $V_S$, which can be seen as follows: let
us for simplicity set $q=0$ for the time being and ask whether the
total potential
\begin{eqnarray}
V=V_I+V_S\,,
\end{eqnarray}
which is plotted in Fig.~\ref{Fig1}, exhibits a minimum. Expanding $V$ around $\varphi=0$ yields
\begin{eqnarray}
V\approx V_0\tilde{\alpha}\varphi+\frac{\tilde{N}}{a^3}\left|\beta\varphi\right|\,.
\end{eqnarray}
In order to stabilize the radion at $\varphi=0$ we need
\begin{eqnarray}
\frac{\tilde{N}}{a^3V_0}\gg\frac{\tilde{\alpha}}{\beta}=\mathcal{O}(1)\,, \label{cond1stab}
\end{eqnarray}
so that the stabilizing potential $V_S$ is able to prevent the
collapse of the internal dimensions due to the inter brane potential
$V_I$. Since the universe expanded roughly another $60$ e-foldings
after inflation until today, we would need
\begin{eqnarray}
\frac{\tilde{N}}{V_0}\gg e^{180}
\end{eqnarray}
if we want a stable radion at late times, which is clearly an
unreasonable condition. This problem is a simple reflection of the
fact that the inter brane potential does not redshift, whereas the
string gas redshifts like matter. Henceforth, it is not surprising
that the attractive force between the branes wins in the long run.
Notice that the same reason makes this type of stabilization
incompatible with the presence of a cosmological constant
\cite{Ferrer:2005hr}.

If one insists on achieving stabilization via a string gas, there
must be a mechanism present that cancels out $V_I$; luckily, such a
mechanism seems possible in our scenario: once the branes associated
with the orbifold fixed planes approach each other within the string
scale they could annihilate via tachyon decay
\cite{Sen:2002nu,Sen:2002in,Mukhopadhyay:2002en,Rey:2003xs,Chen:2002fp,LLM,Gutperle:2004be}.
%%%%%%%%%%%%%%%%% NEW
It should be noted that the universe itself does not go through a
singularity: the radion gets stabilized at the self dual radius so
that there is no big crunch.
%%%%%%%%%%%%%%%%% end NEW

What is more, one can imagine that this decay contributes to
pre-heating, similar to the mechanism employed in the
cyclic/ekpyrotic scenario
%%%%%%%%%%%%%%%%% NEW
\footnote{Note that the cyclic scenario includes a singular
collision of branes pinned to orbifold fixed planes, which is not
what we are dealing with here: the branes in the scenario at hand
come close to each other (within string length), but do not actually
collide.}
%%%%%%%%%%%%%%%%% end NEW
 or in more recent realizations of brane
inflation as in the KKLMMT proposal
\cite{Kachru:2003sx,Kofman:2005yz,Chen:2006ni}. However, this
mechanism comes with a price: the potential over-production of
relics like cosmic strings. If too many of these un-observed relics
are produced, the model at hand would be ruled out
\cite{Communication_SW_HF} \footnote{ One way to avoid the defect
overabundance problem is to enhance the symmetry which is broken
during the annihilation; this can be achieved by having several overlapping
branes instead of just one \cite{Dasgupta:2004dw,Chen:2005ae}. %We thank R. Brandenberger for pointing out this possibility.
 }.
Hence, we shall assume that a mechanism
to cancel the inter brane potential exists without producing too
many relics.

Since the redshifting of the string gas potential can potentially
spoil stabilization at late times, it is a concern whether this
redshifting will also spoil standard pre-heating methods or
leave them unaffected. Thus, we will address this question in the
next subsection.

\subsection{Pre-heating \label{pre-heating}}
Assuming that the inter brane potential cancels via some unspecified
mechanism near the self dual radius, the complete potential for the
radion is provided by the string gas alone, that is
\begin{eqnarray}
V=\frac{N}{a^3}e^{-\frac{d}{2}\beta\varphi}\sqrt{\frac{q^2}{a^2}+\sinh^2(\beta\varphi)}\,. \label{potreheat}
\end{eqnarray}
We would now like to address the question whether the standard
theory of pre-heating after inflation can be applied. The novel
feature in our model is the dependence of the potential on the scale
factor $a$. If one could neglect this feature, pre-heating would
progress as usual, see e.g.
\cite{Traschen:1990sw,Shtanov:1994ce,Kofman:1994rk,Kofman:1997yn,Greene:1997fu,Felder:2000hj}
for a sample of the extensive literature on the subject.

As a first estimate we can compare the rate at which the potential
changes with the Hubble factor. As we shall see below in
(\ref{rateofchange}), both quantities are of the same order. Hence
we expect any effects due to the redshifting of the potential to be
of the same magnitude as those directly caused by the expansion of
the universe. As a consequence, whenever the Hubble expansion needs
to be included, e.g. in the case of stochastic pre-heating
\cite{Kofman:1997yn} (broad parametric resonance in an expanding
universe), one should also include the time dependence of the
potential.

In order to examine more carefully whether the redshifting of an
inflaton potential can be neglected under the assumption that the
expansion of the universe itself is unimportant, we will focus on a
specific toy model for pre-heating
\cite{Kofman:1997yn,Greene:1997fu}:  narrow parametric resonance
\cite{Traschen:1990sw,Shtanov:1994ce}. It should be noted that
narrow or broad parametric resonances will not be viable reheating
mechanisms if the inflaton is identified with the radion (as in our
case), since the couplings between the radion and other
matter-fields are heavily suppressed
%%%%%%%%%%%%% New
\footnote{Nevertheless, there are possibilities to enhance
suppressed reheating channels by considering large vacuum
expectation values of scalar matter fields after inflation, see e.g.
\cite{Allahverdi:2004ge}.}.
%%%%%%%%%%%%% end NEW
Nevertheless, we will focus on narrow resonance as an instructive
example, since the mechanism is quite simple and very sensitive to
changes in the shape of $V$: any change in the potential during the
time-scale of pre-heating will cause the center of the resonance
band to shift. If this shift is larger than the width of the
resonance band, modes would not stay within the band long enough to
get reasonably amplified. But if the shift is small compared to the
width, narrow resonance will commence in the usual way. As we shall
see below, the latter is the case so that there are no new effects
and/or constraints due to the redshifting potential.

To study pre-heating, let us first expand the potential around the
minimum of the potential at $\varphi_{min}=:\sigma$ and thereafter
couple the radion to a scalar matter field $\chi$. At this first
stage we neglect the expansion of the universe so that
$n:=N/a^3\approx const$ and $l^2:=q^2/a^2\approx const$. The minimum
of (\ref{potreheat}) can be found at
\begin{eqnarray}
\sigma (l) &=& \frac{1}{2\beta} \ln \left( -3l^2+\frac{3}{2}-\frac{\sqrt{36(l^4-l^2)+1}}{2} \right)\,,
\end{eqnarray}
Where we used $d=6$. Note that a minimum only exists for
\begin{eqnarray}
0\leq
l<\tilde{l} \label{cond2stab}
\end{eqnarray}
 with $\tilde{l}:=(\sqrt{12}-\sqrt{6})/6$, leading to
$0\leq \sigma<\ln(2)/(4\beta)$. Expanding the potential around
$\sigma$ leads to
\begin{eqnarray}
V&\approx&\tilde{V}_0+\frac{m^2}{2}\phi^2\,,
\end{eqnarray}
where we used a shifted inflaton  $\phi:=\varphi-\sigma$ and
\begin{eqnarray}
\tilde{V}_0=n e^{-3\beta\sigma} \sqrt{l^2+\sinh^2(\beta\sigma)}\,,
\end{eqnarray}
as well as
\begin{eqnarray}
m\approx\beta \sqrt{\frac{n}{l}}-\beta\frac{63}{4}\sqrt{n}l^{3/2} \,, \label{mofl}
\end{eqnarray}
where we expanded $m$ around $l=0$. Note that $m(l=\tilde{l})\equiv
0$ exactly, so that any value of $m$ can be achieved by
appropriately tuning $l$. We will not need the cumbersome exact
expression for $m$ in the following, hence we shall omit it.

At this point we should step back for a second and estimate the rate
of change of the potential. Using $l\propto 1/a$ and $q\propto1/a^3$
we arrive at
\begin{eqnarray}
\frac{\dot{m}}{m}\approx - H \,,\label{rateofchange}
\end{eqnarray}
where we only kept the leading order term in (\ref{mofl}). Hence, we
naively expect that the expansion of the universe and the
redshifting of the potential lead to comparable effects. This
estimate can be made more concrete at the level of the toy model of
narrow parametric resonance: if we couple the radion to a scalar
matter field via $V_{int}=-g^2\phi^2\chi^2$, the system will be in
the regime of narrow resonance if $g\Phi\ll\sigma\ll m$ holds, where
$\Phi(t)$ is the amplitude of the oscillating inflaton
\cite{Kofman:1997yn}. This condition can be satisfied if we are free
to tune $g$ and $m$  appropriately. Following the analysis of
\cite{Kofman:1997yn} closely, we find the first resonance band of
the resulting Mathieu-equation for $\chi_k$ at the wave-number
$k\approx m/2$ with a width of $\Delta k\approx \tilde{q} m/2$ where
$\tilde{q}:=4g^2\sigma\Phi/m^2\ll1$.

Since parametric resonance usually commences during the first few
oscillations of $\phi$ around its minimum, the characteristic
time-scale is given by the period of these oscillations $T=2\pi/m$.

Turning on the expansion of the universe yields the requirement
\begin{eqnarray}
H\ll\tilde{q}^2m\,,\label{con1H}
\end{eqnarray}
in order for narrow resonance to take place \footnote{Notice that
$g$ is expected to be small in our model. As a consequence,
condition (\ref{con1H}) is not satisfied and pre-heating will not
progress in the regime of narrow parametric resonance.}, otherwise
modes would leave the resonance band too fast \cite{Kofman:1997yn}.
Given that inequality, we can give an upper bound on the change of
the scale factor $a$ within the period $T$: because the scale factor
makes a transition from an inflating one to a solution for a
radiation dominated universe during pre-heating, we can use the
inflationary solution as an upper bound for the change in $a$, that
is
\begin{eqnarray}
\frac{a(t_0+T)}{a(t_0)}\approx e^{2\epsilon\pi\tilde{q}^2}
\end{eqnarray}
where we used $H\approx\epsilon\tilde{q}^2m$ with $\epsilon\ll1$.
This results in a change of the potential's shape via a change in
the inflaton mass
\begin{eqnarray}
m(t_0+T)&\approx&\beta \sqrt{\frac{n}{l}}\\
&\approx& m(t_0)e^{-\epsilon2\pi\tilde{q}^2}\\
&\approx& m(t_0)(1-\epsilon2\pi\tilde{q}^2)\,,
\end{eqnarray}
where we only kept the leading order term in $l$ from (\ref{mofl}),
plugged in $l\propto1/a$ as well as $n\propto1/a^3$ and expanded
around $\tilde{q}=0$. Since the position of the first resonance band
is located at $k=m/2$, we see that the shift of its position is
given by $D_k=\epsilon\pi\tilde{q}^2m(t_0)$. This shift has to be
compared with the width of the band $\Delta k\approx \tilde{q}
m(t_0)/2$. We immediately see that $D_k\ll\Delta k$ and henceforth,
we can safely ignore the slight change of the radion potential.

\subsection{Discussion}
We saw in the previous section that the time dependence of the
inflaton potential does not interfere much with the process of
pre-heating. We estimated the effect on the toy model of narrow
parametric resonance, because this pre-heating mechanism is most
sensitive to changes in the mass of the inflaton. We found that new
effects due to the redshifting potential are comparable to the ones
already present due to the expanding universe.

Hence, we expect no novel features during pre-heating if a string
gas supplies the stabilizing potential for the inflaton, and the
standard theory of pre-heating can be applied (we refer the reader
to
\cite{Traschen:1990sw,Shtanov:1994ce,Kofman:1994rk,Kofman:1997yn,Greene:1997fu}
and follow up papers for the relevant literature). However, whenever
the expansion of the universe itself is crucial, one should also
consider the redshifting of the potential; for example, in the case
of stochastic resonance \cite{Kofman:1997yn} the Hubble expansion
causes a mode to scan many resonance bands during a single
oscillation of the inflaton. Naturally, including the redshifting of
the potential will add to this effect, since the resonance bands
themselves shift, just as in the case of narrow resonance we
examined in the previous section.

There is another issue worth stressing again: since the inflaton is
identified with the radion in our setup, its couplings to matter
fields are heavily suppressed. As a consequence, we do not expect
parametric resonance to be the leading pre-heating channel
%%%%%%%%%%% NEW
(see however \cite{Allahverdi:2004ge} for the possibility of \emph{enhanced pre-heating}),
%%%%%%%%%%% end NEW
but
instead tachyonic pre-heating (see e.g.
\cite{Greene:1997ge,Felder:2000hj,Dufaux:2006ee} and references
therein), which occurs in case of a negative effective mass term for
the matter field. This effect was used to address the moduli problem
in \cite{Shuhmaher:2005mf} and warrants further study
\cite{Shu_prep}.

Yet another possibility to reheat the universe could be provided by
the annihilation of the boundary branes via tachyon decay
\cite{Sen:2002nu,Sen:2002in,Mukhopadhyay:2002en,Rey:2003xs,Chen:2002fp,LLM,Gutperle:2004be}
once the branes come close to each other. A potential hinderance
could be an over-production of relics such as cosmic strings. It
seems possible to avoid this problem in certain
circumstances \cite{Dasgupta:2004dw,Chen:2005ae}, but we postpone a
study of this interesting possibility to a future publication, since
it is beyond the scope of this article.

Last but not least, since the production of the stabilizing string
gas will overlap with the early stages of pre-heating, one should
discuss both processes in a unified treatment.

\section{Conclusions}

In this article, we examined observational consequences of the
recently proposed emerging brane inflation model. After reviewing
the aforementioned model, observational parameters were computed
within the slow roll regime, once and foremost the scalar spectral
index $n_s=0.9659^{+0.0049}_{-0.0052}$. This index is a generic
prediction of emerging brane inflation, independent of model
specific details and in excellent agreement with recent constraints
of WMAP3 and SDSS. Furthermore, based one the COBE normalization
we were able derive a bound onto the fundamental string scale, (\ref{string_coupling}).

Thereafter, we examined the consequences of a redshifting string gas
potential, which arises at the end of inflation. Even though the
radion/inflaton can initially be stabilized, the mechanism fails at
late times as long as there is a contribution to the effective
potential that does not redshift, like a cosmological constant or a
remaining interbrane potential. Consequently, a mechanism to cancel out all
such contributions needs to be found in order for the model to work.

Related to this mechanism, we encountered another potential problem:
since the interaction of boundary branes is responsible for
inflation, but branes have to be absent at late times in order to
keep extra dimensions stable, we concluded that they had to
annihilate after inflation. During this annihilation, which could in
principle be responsible for pre-heating, relics like cosmic strings
are expected to be produced. Mechanisms to avoid an overproduction
of said relicts are conceivable, but warrant further study.

Concerned that pre-heating after inflation might also get disrupted
via the time dependence of the potential, we focused on narrow
parametric resonance as a toy model for pre-heating to estimate the
magnitude of new effects: we find that new effects are comparable to
those originating directly from the expansion of the universe.
Henceforth, we concluded that the standard machinery of pre-heating
can be applied to the model at hand, but the time dependence of the
potential needs to be incorporated if expansion effects are crucial
for pre-heating, as is the case in e.g. stochastic resonance. Since
the annihilation of boundary branes and the production of the
stabilizing string gas occurs during the early stages of
pre-heating, one should incorporate these effects in a detailed
study of pre-heating.

To summarize, the proposal of emerging brane inflation is a viable
realization of inflation, if the
potential problems associated with the annihilation of branes after
inflation can be overcome.
%%%%%%%%%%%%%%%%%%%%%%%%%%%%%%%%%%%%

%%%%%%%%%%%%%%%%%%%%%%%%%%%%%%%%%%%%
\begin{acknowledgments}

We would like to thank Robert Brandenberger for many helpful
discussions as well as Diana Battefeld, Cliff Burgess, Jerome Martin and Scott
Watson for comments on the draft. T.B. would like to acknowledge the hospitality of Yale
University and McGill University. N.S. would like to acknowledge the
hospitality of the Perimeter Institute and to thank Justin Khoury
for useful comments regarding dynamics of the scale factors as seen
in the Einstein frame.
% ackn. of Fellowships is frowned upon in APS.
% N.S. would like to acknowledge support from a
% Carl Reinhardt McGill Major Fellowship.

\end{acknowledgments}
%%%%%%%%%%%%%%%%%%%%%%%%%%%%%%%%%%%%

\end{document}